\documentclass[aps,prd,floatfix,preprint,nofootinbib]{revtex4-1}
\usepackage{graphicx}
\usepackage{epic}
\usepackage{eepic}
\usepackage{latexsym}
\usepackage{amssymb,amsmath}

\newcommand{\eq}[1]{(\ref{#1})}
\newcommand{\be}{\begin{equation}}
\newcommand{\ee}{\end{equation}}
\newcommand{\bea}{\begin{eqnarray}}
\newcommand{\eea}{\end{eqnarray}}

\newcommand{\hs}[1]{\hspace{#1 mm}}

\newcommand{\df}{\dot{\phi}}
\newcommand{\zo}{\zeta_k^{(0)}}
\newcommand{\mt}{\tilde{M}_p}
\newcommand{\lf}{\left<}
\newcommand{\rg}{\right>}

\def\a{\alpha}
\def\b{\beta}

\def\d{\delta}

\def\e{\epsilon}

\def\f{\phi}
\def\fr{\frac}
\def\F{\Phi}
\def\vf{\varphi}

\def\l{\lambda}

\def\m{\mu}

\def\s{\sigma}

\def\th{\theta}

\def\vth{\vartheta}

\def\z{\zeta}

\def\o{\omega}

\def\del{\partial}

\let\bm=\bibitem
\def\nn{\nonumber}

\begin{document}

\title{Loops in reheating and cosmological perturbations} 

\author{Ali Kaya}
\email[]{ali.kaya@boun.edu.tr}
\affiliation{Bo\~{g}azi\c{c}i University, Department of Physics, 34342, Bebek, \.{I}stanbul, Turkey}

\date{\today}

\begin{abstract}

We show that in scalar field inflationary models, the loop corrections in reheating corresponding to the decay of the inflaton can cause nontrivial superhorizon evolution of the curvature perturbation. The effect turns out to be prominent when the decay occurs via parametric resonance, even indicating the breakdown of the perturbation theory, as we demonstrate in a specific model.

\end{abstract}

\maketitle

\section{Introduction}

The conservation of the curvature perturbation $\zeta$ on superhorizon scales is an important feature for the inflationary predictions to hold. According to the standard picture, the Fourier transformed curvature perturbation $\zeta_k$ becomes classical as it crosses the horizon during inflation and its amplitude freezes out until it reenters the horizon at a later epoch. The constancy of $\zeta$ is shown to be maintained during inflation in single field models including all loop corrections \cite{loop1, loop2} (see also \cite{ekw}).  However, in the presence of the entropy perturbations, which  exist generically in multi-field models, this property is well-known to break down (see e.g. \cite{mfb,ent};  the loops of  bosonic and fermionic entropy perturbations are known to yield series IR divergencies, see e.g.  \cite{ir1,ir2} ). 

In scalar models, the nearly exponential expansion is usually followed by a phase of coherent inflaton oscillations and the Universe is reheated by the decay of the inflaton. This decay process can be formulated as quantum particle production in a time dependent background and in some cases it may happen in the parametric resonance regime, which is called preheating \cite{reh1,reh2,reh3,reh4}. As pointed out in \cite{mph1,mph2,mph3}, the metric perturbations can be amplified on superhorizon scales during preheating. Although the original proposal was shown to be inefficient due to the suppression of the reheating scalar modes during inflation \cite{mpk1,mpk2,mpk3},  this problem is shown to be absent in some specific models \cite{mph4,mph5,mph6}. Remarkably,  the superhorizon amplification can be achieved without violating the causality due to the coherency of the inflaton oscillations \cite{mph1}.

In this paper, we consider the well-known chaotic $m^2\f^2$ model with the interaction $\f^2\chi^2$ involving a massless scalar $\chi$ responsible for the inflaton decay and calculate 1-loop corrections to the $\zeta$-$\zeta$ correlation function during reheating. Assuming that the standard picture holds,  this model is actually ruled out by Planck with a 95\% confidence level \cite{pl}, but  our aim is to see whether loop corrections are significant and there are various good reasons to study this problem:  First, as pointed above, at least in some preheating models the superhorizon metric fluctuations are known to be affected by the inflaton decay, even at the linearized level. Second, when reheating occurs by the decay of the inflaton to another scalar, the constancy of $\zeta$ is not guaranteed since entropy perturbations exist in two-field models. Third, the mode functions of the reheating scalar $\chi$ enter in the loops of the $\zeta$-$\zeta$ correlation function and thus if the decay occurs in the parametric resonance regime the loop contributions can be greatly enhanced. 

According to the standard lore, $\zeta_k$ becomes classical as it crosses the horizon. Therefore, the subsequent loop effects can be thought to give quantum corrections to this classical configuration, similar to the usual treatment of solitons.\footnote{I would like to thank Robert Brandenberger for suggesting this interpretation.} By causality, the loop effects are expected to be effective on superhorizon scales as long as the inflaton oscillations are coherent \cite{mph1} (the coherence of the oscillations can be lost earlier than expected \cite{coh1,coh2}). Indeed, unless some form of approximation is employed there should not arise any violation of causality since the equations are relativistic by construction and causality is guaranteed to be preserved.  An important technical complication in this computation is that the $\zeta$-gauge, which is defined by setting the inflaton perturbation to zero, becomes ill defined when the inflaton velocity is zero at the oscillation turning points. Therefore, it turns out to be  more convenient to carry out the calculation in the $\zeta=0$ gauge (although the $\zeta$ variable can still be used with care, see the appendix). One may switch to the $\zeta$-gauge at any suitable time, e.g. when  the coherence of the inflaton oscillations is lost, after which the loops are expected to affect the local physics only and $\zeta$ can be assumed to be conserved on superhorizon scales. This will be the main strategy of our computations.  

Normally, in a field theory computation one would expect the loop corrections to be small since the perturbation theory is usually applicable. As we will see, this may come out to be wrong in the presence of parametric resonance effects. It turns out that the coupling constant in the theory, which is taken to be small for the applicability of perturbation theory, is actually replaced by an effective one that is dressed by the growing mode functions. Moreover, there are quite large scales in the problem, i.e. the scale of the instability bands of the growing mode functions and  the background value of the inflaton amplitude, which are a few orders of magnitude smaller than the Planck scale, and these enter in the loop expressions in a nontrivial way. As a result, the quantum effects are expected to enlarge during preheating. Indeed, in the $\l \f^4$ model with the interaction $\f^2\chi^2$, which has a very similar parametric resonance structure with the $m^2\f^2$ model, the preheating era has been shown to yield  non-gaussianities as large as $f_{NL} >{\cal O}(1000)$ \cite{yeni}.  Our findings in this paper are consistent with such estimates. 

\section{The model and loop corrections in reheating}

We consider the following chaotic model with the potential
\be\label{pot} 
V(\phi,\chi)=\fr12 m^2\phi^2+\fr12 g^2 \phi^2\chi^2,
\ee
where $\phi$ is the inflaton and $\chi$ is the scalar that reheats the universe as a result of $\phi\to\chi$ decay. The background metric is taken to be the usual Friedman-Robertson-Walker one 
\be
ds^2=-dt^2+a^2(dx^2+dy^2+dz^2).
\ee
During inflation $\chi=0$ and $\phi>\mt$, where $\mt$ denotes the Planck mass,\footnote{We define $M_p$ and $\mt$ to be the reduced and the usual Planck masses, where $M_p^2=8\pi G/3$ and $\mt=G^2$.}  and inflation ends around $\phi\sim \mt /20$ (see e.g. \cite{reh4}). 

Assuming 
\be
m\gg H,
\ee
where $H$ denotes the Hubble parameter $H=\dot{a}/a$, the postinflationary evolution of the background during reheating can be determined as 
\be\label{b1}
\phi(t)\simeq\Phi \sin(mt),
\ee
where 
\be \label{fh}
\dot{\Phi}+\fr{3H}{2}\Phi\simeq0
\ee
and 
\be\label{ha}
a\simeq\left(\fr{t}{t_R}\right)^{2/3},\hs{5}H\simeq\fr{2}{3t}.
\ee
The time $t_R$ marks the beginning of reheating and the  initial inflaton amplitude can be taken as $\Phi_0\simeq \mt /20$. The background still has $\chi=0$, which can be assumed until the backreaction effects are set in. 

Taking the metric in the ADM form 
\be\label{admd}
ds^2=-N^2dt^2+h_{ij}(dx^i+N^i dt)(dx^j + N^j dt),
\ee
the perturbations are defined as
\bea
&&h_{ij}=a^2e^{2\zeta}\d_{ij},\nn\\
&&\phi=\phi(t)+\vf,\\
&&\chi=0+\chi.\nn
\eea
The Lapse $N$ and the shift $N^i$ must be determined from the constrained equations \cite{mal} (see \eq{ls} below) and the resulting metric  takes the standard form (see e.g. \cite{luc}). Note that we use the same letter $\chi$ to denote the reheating scalar fluctuations since the background value of $\chi$ vanishes. Under an infinitesimal coordinate transformation $x^\m\to x^\m+k^\m$, the leading order change of a tensor field $T$  is given by the Lie derivative
\be
\d T={\cal L}_{k}T. 
\ee
Thus, the covariance of the field equations under the transformation  $t\to t +k^0$ implies the following gauge freedom for the perturbations $\zeta$, $\vf$ and $\chi$:
\be\label{gt}
\d\zeta=Hk^0,\hs{5}\d\vf=\dot{\phi}k^0,\hs{5}\d\chi=0. 
\ee
Using this freedom it is possible to set $\zeta=0$ or $\vf=0$ provided $H\not=0$ or $\dot{\phi}\not=0$, respectively. In inflationary loop calculations, one usually employs the $\vf=0$ gauge since $\zeta$ is taken to be conserved on superhorizon scales and thus it is enough to determine the loop integrals till the first horizon crossing time which sets an upper limit for the time integrals. From \eq{gt}, we see that the $\vf=0$ gauge breaks down when $\dot{\phi}=0$, which occurs repeatedly during reheating. As another manifestation of this issue, the $\zeta$ kinetic term in the $\vf=0$ gauge becomes $(\dot{H}/H^2)\dot{\zeta}^2$, which is degenerate when $\dot{H}=0$, and therefore when $\dot{\phi}=0$. As a result, in what follows we choose the $\zeta=0$ gauge for the calculation of the loop effects during reheating. It is clear that the superhorizon evolution of $\vf$ must be carefully taken into account in this calculation as we discuss below.\footnote{There are other ways of circumventing the breakdown of the $\vf=0$ gauge. One possibility is to impose it for all times except in some time intervals near the turning points $\dot{\phi}=0$, see the appendix. Alternatively, one may include higher order terms in the tensor transformation expression in the form $\d T={\cal L}_{k}T +\fr12{\cal L}_{k}^2T+...$ so that the gauge can still be imposed.}

We follow \cite{mal} and use the ADM decomposition \eq{admd} to write the action as
\be\label{a}
S=\fr12 \int \sqrt{h} \left[NA+\fr{B}{N}\right],
\ee
where
\be
A=R^{(3)}-2V-h^{ij}\del_i\phi\del_j\f-h^{ij}\del_i\chi\del_j\chi,\nn
\ee
and \be
B=K_{ij}K^{ij}-K^2+(\dot{\f}-N^i\del_i\f)^2+(\dot{\chi}-N^i\del_i\chi)^2.\nn
\ee
As noted in \cite{ekw}, the lapse $N$ can be solved algebraically as $N^2=B/A$, which can be used back in the action to eliminate $N$ completely. However, in the following we will only need first order solutions for $N$ and $N^i$ and thus proceed with the usual treatment. After setting the tensor perturbations to zero, imposing the gauge  
\be\label{zg}
\zeta=0
\ee
and using the background equations for some simplifications,  the lapse and the shift  can be solved as 
\be\label{ls}
N=1+\fr{\dot{\f}}{2H}\vf,\hs{5}N^i=\d^{ij}\del_j\psi,
\ee
where 
\be
\del_i\del^i\psi=-\fr{1}{4H}\left[\fr{2\overline{V}}{H}\df\vf+2\overline{V_\f}\vf+2\overline{V_\chi}\chi+2\df\dot{\vf}
\right].
\ee
Here a subindex denotes a partial derivative and an overline indicates the background value. Comparing with the expressions given in \cite{mal}, we see that the presence of the second scalar $\chi$ does not change the lapse $N$. On the other hand, \eq{pot} shows that $\overline{V_\chi}=0$ in our case. From \eq{gt} the
 Faddeev-Popov determinant of the gauge \eq{zg} can be seen to be trivial. 

Using \eq{ls} in \eq{a}, one may obtain the following quadratic action for $\vf$ and $\chi$: 
\be
S^{(2)}=\int\fr{a^3}{2}\left[\dot{\vf}^2-\fr{1}{a^2}(\del\vf)^2+\dot{\chi}^2-\fr{1}{a^2}(\del\chi)^2-m^2\vf^2-g^2\f^2\chi^2-\fr{m^2\f\df}{H}\vf^2-\fr{\f^2\df^2}{4H^2}\vf^2-\fr{\df^2}{H}\vf\dot{\vf} \right].
\ee 
As usual, the fields can be expanded as 
\bea
\vf=\fr{1}{(2\pi)^{3/2}}\int d^3k\, e^{i\vec{k}.\vec{x}}\,\vf_k(t) a_{\vec{k}}+h.c.\label{m1}\\
\chi=\fr{1}{(2\pi)^{3/2}}\int d^3k\, e^{i\vec{k}.\vec{x}}\,\chi_k(t) \tilde{a}_{\vec{k}}+h.c.\label{m2}
\eea
where the canonical quantization implies the standard commutation relations for the ladder operators provided  the following Wronskian conditions are satisfied
\be\label{w}
\vf_k\dot{\vf}_k^*-\vf_k^*\dot{\vf}_k=\chi_k\dot{\chi}_k^*-\chi_k^*\dot{\chi}_k=\fr{i}{a^3}.
\ee
The mode functions obey
\be
\ddot{\vf}_k+3H\dot{\vf}_k+\left[m^2+\fr{2m^2\f\df}{H}+3\df^2-\fr{\df^4}{2H^2}+\fr{k^2}{a^2}\right]\vf_k=0\label{f}
\ee
and  
\be\label{c}
\ddot{\chi}_k+3H\dot{\chi}_k+\left[g^2\f^2+\fr{k^2}{a^2}\right]\chi_k=0.
\ee
The initial conditions for the mode functions must be imposed during inflation when the wavelengths are well within the horizon (naturally referring to the Bunch-Davies vacuum). As can be seen from \eq{c}, the $\chi$-field becomes very massive during inflation since $\f\simeq M_p$ and thus the amplitude $\chi_k$ is suppressed like $a^{-3/2}$, which  prevents the amplification of the curvature perturbation at the linear level as discussed in \cite{mpk1,mpk2,mpk3}. 

To determine the evolution of $\vf_k$ on superhorizon scales, for which the $k^2/a^2$ term in \eq{f} is negligible, we first note that $\vf_k=\df/H$ is a solution of \eq{f} with $k=0$. This solution can be obtained by a gauge transformation from a constant $\z_k$ configuration in the $\vf=0$ gauge. Therefore, the coefficient of this solution can be identified with the constant superhorizon value of the curvature perturbation, which we denote as $\zo$. The second linearly independent solution can be found using the Wronskian to be $\vf_k\simeq \df f/H$, where 
\be\label{df}
\fr{df}{dt}=\fr{H^2}{a^3\df^2}.
\ee
Therefore, on superhorizon scales one finds
\be\label{sh}
\vf_k\simeq \fr{\df}{H}\left[\zo+c_k f(t)\right],  
\ee
where $c_k$ is a complex constant. During inflation $f\sim 1/a^3$ and thus \eq{sh} contains  the  ``constant" and the decaying pieces as expected. The normalization \eq{w} implies
\be\label{n}
\zo c_k^* -\zo{}^* c_k=i.
\ee
In principle  both $\zo$ and $c_k$ can be determined from the evolution equation \eq{f} provided that  the initial conditions are given. In finding the solution \eq{sh} we have not employed any specific property of the background and thus it is also valid during reheating, which can be verified directly from \eq{f} by using the background field equations. Although $df/dt$ diverges at times $t$ obeying $\df(t)=0$, the solution \eq{sh} is well defined since around such a time $f\sim1/\df$ and thus $\df f$ is well behaved (see \eq{fs}). 

As emphasized above, our strategy is to carry out the loop calculation in the $\z=0$ gauge until the coherence of the inflaton oscillations is lost and then switch to the $\vf=0$ gauge to determine the $\z$ correlation function. Using \eq{gt}, the set $\z=0$ and $\vf(t,\vec{x})$ can be transformed into $\vf=0$ and $\z(t,\vec{x})$ by choosing $k^0=-\vf(t,\vec{x})/\df(t)+{\cal O}(\vf^2)$, which would give
\be\label{tr}
\z(t,\vec{x})=-\fr{H}{\df}\vf(t,\vec{x})+{\cal O}(\vf^2),
\ee
where ${\cal O}(\vf^2)$ terms necessarily appear in the full gauge transformation. When one changes the variable from $\vf$ to $\zeta$ in the action, the higher order nonlinear terms in \eq{tr} give interactions with more powers of the field variables and thus supposedly are less important than the linear term. Note that since the background $\chi$ field vanishes, $\chi$ fluctuation does not change under a coordinate transformation. Viewing as an operator equation,  \eq{tr} can be used to convert $\vf$-spectrum to $\z$-spectrum which implies
\be\label{r}
\lf \z(t,\vec{x})\z(t,\vec{y})\rg=\fr{H^2}{\df^2}\lf \vf(t,\vec{x})\vf(t,\vec{y})\rg+...
\ee
where the dots denote the contributions of the higher order terms noted above. Defining the power spectrum $P_k^\vth(t)$ corresponding to a scalar field $\vth$ by 
\be\label{ps}
\lf \vth(t,\vec{x})\vth(t,\vec{y})\rg=\fr{1}{(2\pi)^{3}}\int d^3k \,e^{i\vec{k}.(\vec{x}-\vec{y})}\,P_k^\vth(t),
\ee
\eq{r} gives
\be\label{tr2}
P^\z_k(t)\simeq\fr{H^2}{\df^2}P_k^\vf(t). 
\ee
Using \eq{m1}, \eq{sh}, \eq{ps} and \eq{tr2} one may obtain the tree-level result (note that  the decaying solution in \eq{sh} is negligible here)
\be
P_k^{\z(0)}(t)=|\zo|^2
\ee
and our aim is to use the same formulas in the loop calculations during reheating. As we will see $P_k^\vf$ turns out to have enough powers of $\df$ to make \eq{tr2} well defined for $P_k^\z$ even when $\df=0$. However, it is enough for us to evaluate \eq{tr2} at the time $t_f$, just before the coherence of the inflaton oscillations is lost (i.e. at the end of the first stage of preheating), which can be chosen so that $\df(t_f)\not=0$. 

The expansion of the action \eq{a} around the background solution gives various interaction terms but the effect of the reheating loops on the power spectrum of $\z$ can be seen most straightforwardly from the interaction potential \eq{pot} that gives a quartic coupling involving $\vf$ and $\chi$. The corresponding interaction Hamiltonian is given by 
\be\label{i1}
H_I(t)=\fr12 g^2\, a(t)^3\int \, d^3z\,\vf(t,\vec{z})^2\chi(t,\vec{z})^2.
\ee
For any given operator $O$, the in-in formalism gives the following perturbative expansion for the vacuum expectation value \cite{w0} 
\be\label{inp}
\left< O(t)\right>=\sum_{N=0}^{\infty} i^N \int_{t_i}^t dt_N\int_{t_i}^{t_N}dt_{N-1}...\int_{t_i}^{t_2}dt_1\left< [H_I(t_1),[H_I(t_2),...[H_I(t_N),O(t)]...]\right>.
\ee
Using \eq{inp} for the operator $\vf(t,\vec{x})\vf(t,\vec{y})$ with $N=1$, one can evaluate the contribution of the 1-loop diagram of Fig.1a to the power spectrum $P_k^\vf$ as 
\be
P_k^\vf(t)^{(1)}=ig^2\int_{t_i}^t dt'a(t')^3\lf\chi^2(t')\rg\left[\vf_k(t')^2\vf_k^*(t)^2-\vf_k^*(t')^2\vf_k(t)^2\right].\label{l1}
\ee
In deriving this expression, one must first evaluate the following commutator that arises from \eq{inp}
\be
\left<\chi^2(t',\vec{z})[\vf(t',\vec{z})^2,\vf(t,\vec{x})\vf(t,\vec{y})]\right>
\ee
by using the expansion \eq{m1} and then utilize \eq{ps} to obtain the power spectrum in momentum space. The initial conditions for the mode functions are imposed at $t_i$ referring to the vacuum of the theory. In principle, $t_i$ must correspond to the beginning of inflation but one usually lets $t_i\to-\infty$. Although we write them in the same integral in \eq{l1}, the $i\e$ prescriptions for the two terms in the square brackets are different, which is important for the convergence of the integral as $t_i\to-\infty$. This technical complication will not be important for us since we are interested in the loop contributions during reheating. For now the time $t$ refers to an arbitrary time during reheating, but we eventually set $t=t_f$, where at $t_f$ the coherence of the inflaton oscillations is lost, as defined above. 

\begin{figure}
\centerline{
\includegraphics[width=9cm]{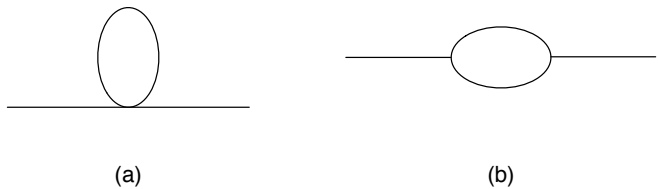}}
\caption{The 1-loop graphs arising from the potential \eq{pot} and contributing to the $\zeta$-$\zeta$ correlation function during reheating. The same graph gives various $\pm$ vertex contributions, which are not shown explicitly.} 
\label{fig1}
\end{figure}

The terms in the square brackets in \eq{l1} come from a commutator and as shown in \cite{w} they decay like $1/a^3$ canceling the $a^3$ factor arising from the spatial volume measure. Therefore, \eq{l1} is negligible during inflation since $\lf \chi^2\rg $ cannot take large values. In the reheating stage, the mode function $\vf_k$ takes the asymptotic form \eq{sh}. Using \eq{sh} and \eq{n} in \eq{l1} one easily finds
\be
P_k^\vf(t)^{(1)}\simeq 2g^2\fr{\df(t)^2}{H(t)^2}|\zo|^2\int_{t_R}^t dt'a(t')^3\lf \chi^2(t')\rg\fr{\df(t')^2}{H(t')^2}\left[f(t')-f(t)\right],
\ee
where $t_R$ denotes the beginning of reheating. Converting the power  spectrum from \eq{tr2} we obtain
\be\label{r1}
P_k^\z(t)^{(1)}\simeq P_k^{\z(0)}\, 2g^2\int_{t_R}^t dt'a(t')^3\lf \chi^2(t')\rg \fr{\df(t')^2}{H(t')^2}\left[f(t')-f(t)\right].
\ee
This result clearly shows the superhorizon evolution of the power spectrum as $\lf \chi^2\rg $ gets larger in time. Eq. \eq{r1} corrects the tree-level amplitude but does not change the index of the spectrum since there is no dependence on the wave number $k$. Because \eq{r1} depends on the difference of two $f$-functions, there is no need to fix the integration constant that arises from  \eq{df}. 

Let us now estimate the correction \eq{r1}. As discussed in \cite{reh4}, $\lf \chi^2(t')\rg$ increases exponentially in time, which can be described by an effective index $\m$ so that $\lf \chi^2(t')\rg \sim \exp(2\m mt)$. This shows that  after each oscillation $\lf \chi^2(t')\rg$ is enlarged roughly by $\exp(4\pi\m)$ times. In \cite{reh4}, the index $\m$ is determined numerically for various cases. For instance, for $g=10^{-2}$ and $m=10^{-6}\mt$ one has $\m\simeq 0.13$, which is a typical value for $\m$. In that case the first stage of preheating ends after 11 oscillations and  $\lf \chi^2(t')\rg$ is enlarged by $\exp(4\pi\m)\simeq 5$ times after each period. Since the largest  contribution to \eq{r1} comes from the last oscillation in which  $\lf \chi^2(t')\rg$ reaches its maximum value, below we estimate \eq{r1} for this last period  (of course one may keep in mind that the previous cycle gives $1/5$ of this maximum value for $g=10^{-2}$ and $m=10^{-6}\mt$). 

Using \eq{b1} in \eq{df} one finds
\be
\fr{df}{dt}=\fr{H^2}{a^3(\Phi m\cos(mt)+\dot{\F}\sin(mt))^2}.
\ee
Since $H$, $a$, $\F$ and $\dot{\F}$ are slowly varying compared to $\sin(mt)$ and $\cos(mt)$, one can approximately integrate this equation by treating them as constants to obtain 
\be\label{fs}
f\simeq\fr{H^2\sin(mt)}{a^3\Phi m^2\df}.
\ee
Note that by \eq{sh}, the actual mode function $\dot{\f}f/H$ decreases like $1/a^3$. One may first think that due to this falloff behavior, $f(t)$ can be neglected compared to $f(t')$ in \eq{r1}. However,  $f$ is also oscillating  and thus one must consider  both terms in \eq{r1}. Let us label the respective corrections by I and II.  Using \eq{fs} for $f(t')$, the first part of the correction becomes
\be
\textrm{I}\simeq 2g^2\int_{t_R}^t dt' \lf \chi^2(t')\rg \fr{\df(t') \sin(mt')}{m^2\F}.
\ee
The $\cos(mt')$ term in $\df(t')$ gives an oscillating factor that is negligible. Using \eq{fh}, one may then find 
\be
\textrm{I}\simeq 3g^2\int_{t_R}^t dt' \lf \chi^2(t')\rg  \fr{H(t')\sin^2(mt')}{m^2}.
\ee
In integrating this expression in the last oscillation cycle, one can treat $\lf \chi^2(t')\rg$ and $H$ as constants to obtain 
\be\label{r2}
\textrm{I}\simeq3\,\pi\, g^2\,\fr{H\lf \chi^2\rg}{m^3},
\ee
where $\lf\chi^2\rg$ denotes the average value of $\lf \chi^2(t')\rg$ and the factor $\pi$ arises 
from the integral of $\sin^2(mt')$ for one period. For the correction II, one may use $\dot{\f}(t')\simeq m\Phi\cos(mt')$ since $f(t)$ does not vary. This gives
\be
\textrm{II}\simeq 2g^2\int_{t_R}^t dt'a(t')^3\lf \chi^2(t')\rg \fr{m^2\Phi^2\cos(mt')^2}{H(t')^2}\fr{H(t)^2\tan(mt)}{a(t)^3\Phi^2m^3},
\ee
where we use \eq{fs} to replace $f(t)$. Again treating the slowly varying factors as constants and using \eq{ha} one may obtain 
\be\label{r22}
\textrm{II}\simeq 2Cg^2 \fr{\lf \chi^2\rg}{m^2},
\ee
where $C$ is the value of the following dimensionless integral
\be\label{ni}
C=\int_{\textrm{last cycle}} (mdt') \cos^2(mt') \left[\fr{t'^4}{t^4}\right]\tan(mt).
\ee
Therefore, the total one loop correction corresponding to Fig. 1a  becomes 
\be\label{ne1}
P_k^\z(t)^{(1)}\simeq g^2\left[\fr{3\pi H}{m^3}+\fr{2C}{m^2}\right] \lf \chi^2\rg P_k^{\z(0)},
\ee
As we see below, $C$ turns out to be an order one number that depends on the parameters of the model. Since $m\gg H$, correction II becomes larger than correction I. 

Another way of estimating the correction \eq{r1} is as follows:\footnote{We thank the anonymous referee for suggesting this alternative.}  Eq. \eq{ha} implies $\dot{H}\simeq -3H^2/2$ and thus  the background field equation $M_p^2\dot{H}=-\fr12 \dot{\f}^2$ gives $|\dot{\f}^2|\simeq 3H^2M_p^2$. Note that $\dot{\f}$ is an oscillating function and the last equality only estimates how the amplitude of $\dot{\f}$ decreases in time. Using this expression in \eq{df}, one may  obtain
\be\label{fa}
|f|\simeq \fr{2}{9a^3}\fr{1}{HM_p^2},
\ee
which fixes the time dependence of the  amplitude of the function $f$. Using \eq{fa} in \eq{r1} one finds
\be
P_k^\z(t)^{(1)}\simeq P_k^{\z(0)}\, 2g^2\int_{t_R}^t dt'\lf \chi^2(t')\rg \fr{m^2\Phi^2\cos(mt')^2}{H(t')^3}\fr{2}{9M_p^2}\left[1-\fr{H(t')a(t')^3}{H(t)a(t)^3}\right]. 
\ee
As pointed out above, to estimate the integral one may focus on the last oscillation cycle that gives the maximum contribution. Once more introducing the average quantities for slowly changing functions and using \eq{ha}, one may obtain
\be\label{ne2}
P_k^\z(t)^{(1)}\simeq P_k^{\z(0)}\, \fr{4}{9}\,\tilde{C}\, \lf \chi^2\rg \fr{g^2m\Phi^2}{H^3M_p^2},
\ee
where 
\be\label{nii}
\tilde{C}=\int_{\textrm{last cycle}} (mdt') \cos^2(mt') \left[\fr{t'}{t}-1\right].
\ee
Again, $\tilde{C}$ turns out to be an order one dimensionless number. In our previous estimation, the contribution of $f(t')$  in \eq{r1} was subleading. This was due to the fact that $f(t')$ is an oscillating function. In \eq{ne2}, this property is overlooked since \eq{fa} only determines the evolution of the amplitude. Utilizing this property in \eq{ne2} gives a modified constant
\be\label{niic}
\tilde{C}=\int_{\textrm{last cycle}} (mdt') \cos^2(mt') \left[\fr{t'}{t}\right].
\ee
We see below that together with this revision the two estimates \eq{ne1} and \eq{ne2} agree with each other. 

Eq. \eq{ne1} or \eq{ne2} give the 1-loop correction to the power spectrum at the end of the first stage of reheating during which inflaton oscillates coherently. To simplify these estimates, we note that in this model the backreaction kicks in when the interaction potential energy $g^2\chi^2\f^2$ catches up the inflaton potential energy $m^2\f^2$. Therefore, the first stage of reheating ends when 
\be\label{cmax}
\lf\chi^2\rg \simeq \fr{m^2}{g^2}.  
\ee
It is known that if the decay occurs in the parametric resonance regime $\lf\chi^2\rg$ can have enormously large values \cite{reh4}, which would give a large loop correction during reheating. For instance, if  $g=10^{-2}$, $m=10^{-6}\mt$, one has $\lf\chi^2\rg\sim 10^{-8} \mt^2$.   In that case $H\simeq 10^{-2}m$ at the end of the first stage of preheating and one has $mt_f\simeq 10^{-2}Ht_f\simeq 200/3$, where we have used \eq{ha}. Numerically integrating \eq{ni} and \eq{niic} from $mt_f-2\pi$ to $mt_f$ gives $|C|\simeq 2.23$ and $|\tilde{C}|\simeq3.02$. Then, the one loop correction can be determined from \eq{ne1} and \eq{ne2} to give 
\be\label{cases}
P_k^{\z(1)}\simeq \begin{cases}4.45\,P_k^{\z(0)},\\4.00\,P_k^{\z(0)},\end{cases}
\ee
respectively. We see that the two estimates fairly agree with each other. In Appendix \ref{ap3}, we numerically integrate \eq{r1} not only in the final period but in the whole first stage of preheating. This gives\footnote{The constants $C$, $\tilde{C}$ and the one in \eq{mc}  are actually negative. However, this is an artifact of perturbation theory, which is similar to the expansion $e^x=1+x+...$ with $x<-1$ that gives a negative number when only the first two terms are kept. Therefore, the results \eq{cases} and \eq{mc} must be understood with this provision.}
 \be\label{mc}
 P_{k\,\,\textrm{num}}^{\z(1)} \simeq  3.38\,P_k^{\z(0)},
\ee
which is consistent with \eq{cases}. The total power spectrum is given by the sum of the tree-level result $P_k^{\z(0)}$ and the 1-loop correction $P_k^{\z(1)}$; therefore the amplitude of the observed power spectrum is changed considerably. 

As pointed out above, in this model the analysis of the linearized field equations show that the amplification of the curvature perturbation on superhorizon scales is not possible due to the suppression of $\chi_k$ modes during inflation \cite{mpk1,mpk2,mpk3}.  This problem is bypassed in the loop  since the effect depends on $\lf \chi^2\rg$, which depends  on the integral of the momentum modes. The loop infinity is also hiding in $\lf\chi^2\rg$, which must be suitably renormalized as discussed in \cite{reh4}. 

The correction \eq{r1} does not change the index of the spectrum; however there are other interactions which can (although the change becomes very small for cosmologically interesting scales, as we will see below). Consider the cubic interaction term that arises from the interaction potential \eq{pot} with one inflaton field is set to its background value. The corresponding interaction Hamiltonian is given by 
\be\label{i2}
H_I(t)= g^2\f(t)\, a(t)^3\int \, d^3x\,\vf(t,\vec{x})\chi(t,\vec{x})^2. 
\ee
The 1-loop correction to the power spectrum $P_k^\vf$ arising from this interaction involves two $H_I$ vertices and the corresponding graph is shown in Fig.1b. In this case, one must use \eq{inp} for the operator $\vf(t,\vec{x})\vf(t,\vec{y})$ with $N=2$ that gives the following nested commutators 
\be
\left<\left[\vf(t_1,\vec{z}_1)\chi^2(t_1,\vec{z}_1),\left[\vf(t_2,\vec{z}_2)\chi^2(t_2,\vec{z}_2),\vf(t,\vec{x})\vf(t,\vec{y})\right]\right] \right>.
\ee
Using the expansions \eq{m1} and \eq{m2}, and applying the commutator identity $[AB,C]=A[B,C]+[A,C]B$,  one may obtain
\bea
&&P_k^\vf(t)^{(1)}=\fr{4g^4}{(2\pi)^3}\int_{t_R}^t dt_1\int_{t_R}^{t_1}dt_2\,\int d^3q\,a(t_1)^3\,
a(t_2)^3 \label{1b} \\
&&\f(t_1)\,\f(t_2)\left[\chi_q(t_1)\chi_{k+q}(t_1)\chi_{q}^*(t_2)\chi_{k+q}^*(t_2)\right] \vf_k(t)\vf_k^*(t_2)\left[\vf_k^*(t)\vf_k(t_1)-\vf_k(t)\vf_k^*(t_1)\right]+c.c. \nn
\eea
where we focus on the reheating contribution by setting the lower limit of the time integrals to $t_R$. Note that $\chi_k$ modes decrease like $a^{-3/2}$  (see \eq{est} below),  i.e. $\chi(t_1)\sim a(t_1)^{-3/2}$ and $\chi(t_2)\sim a(t_2)^{-3/2}$ and these suppressions are compensated by $a(t_1)^3$ and $a(t_2)^3$ factors in \eq{1b}.  On the other hand, the term in the square brackets in the second line also falls like $1/a^3$, which would make the graph completely irrelevant if not canceled out. But, in the parametric resonance regime the instability bands appear in the physical momentum scale. For instance, the first instability band, which gives maximum amplification, is defined in the interval  $q_{phys}\in(0,q_*)$, where $q_*$ is given by \cite{reh4} 
\be
q_*=\sqrt{gm\F},
\ee
where $\F$ denotes the value of $\Phi(t)$ at the end of the first stage of preheating. Therefore, this extra  $1/a^3$ factor can simply change the comoving momentum integral measure $d^3 q$ to the physical one, avoiding the suppression. 

In what follows we estimate \eq{1b} for the modes in the first instability band and restrict the range of the momentum integration variable  to $(0,a q_*)$. We further assume that the subtractions required for the renormalization of the graph are negligible in this interval and as a result we are not bothered by the loop divergence in \eq{1b} (this turns out to be the case as it is shown in \cite{a2}). Using \eq{sh} and \eq{n}, the last two terms in the second line of \eq{1b}  approximately become
\be\label{mm}
i\fr{\df(t)^2\df(t_1)\df(t_2)}{H(t)^2H(t_1)H(t_2)}\left[f(t)-f(t_1)\right] |\zo|^2,
\ee
where we neglect all other terms, which decay at least like $1/a^3$ and become negligible compared to \eq{mm}. Converting the $\vf$-spectrum to the $\z$-spectrum by using \eq{tr2}, one may obtain
\be
P_k^\z(t)^{(1)}\simeq\fr{4ig^4}{(2\pi)^3}\int_{t_R}^t dt_1\int_{t_R}^{t_1}dt_2\,\f(t_1)\,\f(t_2)
\fr{\df(t_1)\df(t_2)}{H(t_1)H(t_2)}\left[f(t)-f(t_1)\right]F\,P_k^{\z(0)},\label{1b2}
\ee
where 
\be\label{F}
F=a(t_1)^3a(t_2)^3\int  d^3q\,\left[\chi_q(t_1)\chi_{k+q}(t_1)\chi_{q}^*(t_2)\chi_{k+q}^*(t_2)-c.c.\right],
\ee
Note that $F$ is purely imaginary and thus \eq{1b2} is real as it should be. 

Next, lest us work out the function $F$. It is known that the evolution of $\chi_q$ given by \eq{c} is nearly adiabatic between successive moments where the inflaton vanishes \cite{reh4} so that it can be written as 
\be\label{est}
\chi_q=\fr{1}{\sqrt{2a^3\o_q}}\left[\a_qe^{-i\int \o_q}+ \b_qe^{i\int \o_q}\right],
\ee
where $\a_q(t_R)=1$, $\b_q(t_R)=0$ and 
\be
\o_q^2=g^2\f^2+\fr{q^2}{a^2}.
\ee
The initial conditions for $\a_q$ and $\b_q$ are fixed by using the initial value of $\chi_q$ in the beginning of  the reheating process, which, up to an irrelevant phase, becomes $\chi_q(t_R)\sim 1/\sqrt{2a^3g\F_0}$ (this is because during inflation $\chi$ becomes  a very massive field with mass $g\F_0$). As the inflaton passes through the potential minimum  $\f=0$, $\chi_q$ changes nonadiabatically and this process can be formulated as the particle creation by parabolic potentials that gives the exponential increase $\b_q = e^{\m_q m t}$, where $\m_q$ is the index describing the exponential growth in a given instability band. Let us introduce the amplitude $A_q$ and the phase $\th_q$ of the mode function as
\be\label{rc}
\chi_q=\fr{1}{\sqrt{2a^3}}\,e^{-i\th_q}A_q.
\ee
As shown in \cite{reh4}, in the broad parametric resonance regime $\th_q$ changes rapidly and thus it behaves like a random variable that  depends sensitively on the initial conditions. To determine $\th_q$,  one may use the Wronskian condition \eq{w}, which implies
\be\label{tq}
\fr{d\th_q}{dt}=\fr{1}{A_q^2}.
\ee 
Using now \eq{rc} in \eq{F} one may find   
\be\label{fson}
F=\fr{i}{2} \int d^3 q A_q(t_1)A_q(t_2) A_{k+q}(t_1)A_{k+q}(t_2)Q
\ee
where 
\be\label{q}
Q=\sin\left[\th_q(t_1)+\th_{k+q}(t_1)-\th_q(t_2)-\th_{k+q}(t_2)\right].
\ee
We see that the phases do not cancel each other since they have different time arguments and this forbids  the cancellation of the leading order term in $F$. 

To proceed, we approximate the momentum integral in \eq{fson}. As pointed out above, we restrict the integral to the first instability band $q\in (0,aq_*)$.  For $k\ll q$, which is true for the cosmologically interesting scales, one may estimate $F$ as 
\be\label{qara2}
F\simeq 2\pi i\,a(t)^3 q_*^3 Q_*  \,A_{q_*}(t_1)^2A_{q_*}(t_2)^2\,\left[1+B\fr{\hat{k}}{q_*}\right]
\ee
where $\hat{k}=k/a$ and $B$ is a dimensionless number of order one that can be fixed by Taylor expanding the integrand in the variable $k$. In the last expression we set $q=aq_*$ since the largest contribution to the integral comes when $q$ runs near $aq_*$. Eq. \eq{qara2} shows that the running of the spectral index is negligible for cosmologically interesting scales since $\hat{k}/q_*$ is an extremely small number. Nevertheless, the scale dependence of the index can be significant for the modes entering the horizon during reheating. It is known that such perturbations can cause formation of primordial black holes during the reheating stage and the modification of the spectral index may change the formation rate and thus the constraints on the inflationary models \cite{bh}.  

As noted above, the amplitude $A_q$ increases exponentially in time, i.e $A_q\sim \exp(\m mt)$. To determine the time dependence of $A_q$ more precisely one may use the following equation 
\be
\lf\chi^2\rg=\fr{1}{(2\pi)^3}\int d^3 q |\chi_q|^2=\fr{1}{2(2\pi a)^3}\int d^3 q A_q^2\simeq \fr{1}{4\pi^2}q_*^3A_{q_*}^2,
\ee
where in the last line we again approximate the integral for $q$ near $aq_*$. From \eq{cmax}, one obtains the maximum value for the amplitude as
\be\label{amax}
A_{q_*}^{max}\simeq \fr{2\pi m}{g q_*^{3/2}}. 
\ee
The time dependence of the amplitude can be fixed from this maximum value as
\be\label{at}
A_{q_*}(t)\simeq \,\fr{2\pi m}{g q_*^{3/2}}\exp(\m m(t-t_f)).
\ee
Recall that $t_f$ denotes the end of the first stage of preheating.  

To complete the estimate, one must finally evaluate the time integrals in \eq{1b2}. As in the previous cases, the largest contribution to \eq{1b2} comes from the last oscillation and  we focus on this cycle. Since after each cycle the amplitude increases by a factor of $\exp(2\pi\m)$, we approximate  
\be\label{aq}
A_{q_*}\simeq \, \fr{2\pi m}{gq_*^{3/2}} \exp(-\pi\m), 
\ee
i.e. we take the amplitude to have its value in the middle of the last period.  In this last cycle, \eq{tq} can be integrated to obtain the time dependence of the phase as
\be\label{faz}
\th_*(t)\simeq \fr{g^2 q_*^3}{4\pi^2m^2}\exp(2\pi\m)\,t.
\ee
As a result, the function $F$ can be approximated as
\be
F_{\textrm{last cycle}}\simeq i(2\pi)^5 a(t)^3\fr{m^4}{g^4q_*^3}\exp(-4\pi \m)Q_*,
\ee
where $Q_*$ must be read from the phases given in \eq{faz}. 

In calculating the contribution of $f(t)$ to \eq{1b2}, the time derivatives on $\df(t_1)$ and $\df(t_2)$ must act on the amplitudes $\F(t_1)$ and $\F(t_2)$, since otherwise one has oscillating functions like $\sin(mt_1)\cos(mt_1)$ whose integrals vanish in one cycle. On the other hand, when one uses \eq{fs} for $f(t_1)$ in \eq{1b2},  $\dot{\f}(t_1)$ factors cancel each other and one only needs to take the derivative of $\F(t_2)$ in calculating $\df(t_2)$ to avoid the oscillating functions. Then, from \eq{fh} it is possible to see that the $f(t_1)$ term becomes ${\cal O} (m/H)$ times larger than the $f(t)$ term in \eq{1b2}. Combining all these factors and treating the slowly changing terms  like $H$ and $\F$ as constants, we obtain the leading order correction as 
\be\label{r3}
P_k^\z(t)^{(1)}\simeq \,P_k^{\z(0)}\,\overline{C}\left[24\pi^2\fr{\F^2H}{q_*^3}\exp(-4\pi \m)\right]\left(1+B\fr{k}{q_*}\right),
\ee
where the dimensionless number $\overline{C}$ is given by
\be
\overline{C}=\int_{\textrm{last cycle}} (mdt_1)\int_{t_1}^t (mdt_2) \sin(mt_1)^2\sin(mt_2)^2 \left[\fr{t}{t_1}\right]^2 Q_*,
\ee
In this last expression, we have ignored the $f(t)$ term and used $a(t)^3/a(t_1)^3=t^2/t_1^2$. Unlike the constants $C$ and $\tilde{C}$ introduced earlier, $\overline{C}$ can be small  due to the presence of the oscillating factor $Q_*$, which can be read from \eq{q}. 
 
Let us now evaluate \eq{r3} again for $g=10^{-2}$ and $m=10^{-6}M_p$. In that case, at the end of the first stage one has $\F\simeq 5\times10^{-3} M_p$ and $H\sim 10^{-2}m$  \cite{reh4}. Moreover, the index is given by $\m\simeq 0.13$. Using these numbers one may calculate the square brackets in \eq{r3} to give  32698 and $\overline{C}$ can be obtained by a numerical integration to yield $\overline{C}\simeq 0.037$. Therefore, \eq{r3} gives  
\be\label{mc2}
P_k^\z(t)^{(1)}\simeq 1227\,P_k^{\z(0)}.
\ee 
Getting a correction much larger  than \eq{cases} indicates that the perturbation theory might be broken down in this model, which would modify the inflationary predictions profoundly. Namely, although the first correction comes with $g^2$, the second one involves $g^4$. Thus, for $g=10^{-2}$ one would expect the second to be smaller about $10^{-4}$ times the first.  Nevertheless, from the interaction potential \eq{pot}, an effective  coupling constant $g_{eff}$ can be defined as 
\be
g_{eff}=g\,e^{2\pi N\mu},
\ee
where the exponential enhancement arises from the $\chi^2$ term in \eq{pot}. For the above example, one may find $g_{eff}\simeq 80$, indicating that the theory might become strongly coupled near the end of preheating. However, to show that the perturbation theory is not applicable one must carefully check all estimates including important numerical factors to make sure that higher order graphs are not suppressed. In particular, the oscillatory character of the time integrals can cause significant reductions, as in \eq{r3}. Moreover, the renormalization of the graphs should be considered in detail, which might modify the amplitudes. In any case, the fact that the resonant $\chi$-modes grow exponentially and more and more $\chi$-mode functions appear at higher orders  clearly threatens the perturbation theory. 

\section{Conclusions}

In this work we show that the loop effects in reheating can meaningfully modify the curvature perturbation power spectrum, which is usually assumed to be conserved on superhorizon scales after horizon exit during inflation. It is known that  the entropy perturbations can change the power spectrum and the effect we have studied is similar in spirit since it involves a second scalar field to which inflaton decays. Nevertheless, the loops arise from nonlinear interactions and thus the two effects are essentially different from each other. Although the chaotic $m^2\f^2$ model is ruled out by Planck with a 95\% confidence level assuming that the standard tree-level results hold \cite{pl}, our findings indicate a possible modification by higher order quantum effects.

The loop corrections to the $\z$-$\z$ correlation function involve time integrals, which are practically limited by the moment of horizon exit  since $\z$ is assumed to be conserved afterword. It is clear that if $\z$ possibly evolves  on superhorizon scales, this upper limit must be extended further. When the limit is pushed to the reheating stage, the $\vf=0$ gauge becomes ill defined at times the inflaton velocity vanishes, and thus it is  more convenient to work with the inflaton fluctuation $\vf$ by imposing the $\z=0$ gauge. In that case, we have observed that the loops involving the reheating scalar $\chi$ become significant at times near the end of the first stage of reheating, just before the backreaction effects become important, i.e.  when the $\chi$ modes are amplified most. As discussed in \cite{mph1}, this superhorizon  influence does not violate causality since the coherently oscillating inflaton background can produce the same effect at different points in space.

In this paper we have only calculated the 1-loop graphs arising from the interaction potential \eq{pot}. There are other types of interactions that came from the solutions of the lapse and shift \eq{ls}. These terms are suppressed during inflation by the slow roll parameters, but this is no longer true during reheating. Indeed, they do not involve a coupling constant, yet in general they are multiplied by the oscillating functions like $\df$ (see \cite{a2}). Thus, it would be interesting to calculate their contribution to the power spectrum. As we saw above, the amplitude corresponding to the graph Fig.1b turned out to be large compared to the tree-level result and therefore it is  crucial to check whether perturbation theory really breaks down during preheating by systematically estimating the contributions of all higher order graphs. 

From our computations, it is possible to understand the reason for the large loop corrections. As emphasized above, one may identify an effective dressed coupling constant $g_{eff}$, which indeed becomes much larger than unity. Therefore, it is not surprising for the correction \eq{mc2} to be larger than \eq{mc}, since the former expression involves $g_{eff}^4$ and the later one has $g_{eff}^2$. On the other hand, there are already very large scales in the problem. The background inflaton amplitude $\F$ is about three orders of magnitude smaller than $M_p$. Similarly, the scale $q_*$ characterizing the instability bands is very large. These scales enter the loop expressions in a nontrivial way, which can increase the magnitude of the corrections. In any case, getting such large corrections during preheating is not surprising. Indeed, in \cite{yeni},  the parametric resonance effects are shown to produce very large non-gaussianities, of the order of 1000, which is consistent with our results. 

\appendix

\section{Using the $\zeta$ variable} 

During reheating $\zeta$ becomes ill defined as a dynamical variable since its kinetic term in the action vanishes when the inflaton velocity becomes zero. This can be seen from the quadratic action
\be\label{a1}
S^{(2)}_\zeta=\int\fr{a^3}{2}\fr{\dot{\f}^2}{H^2}\left[\dot{\zeta}^2-\fr{1}{a^2}(\del\zeta)^2\right].
\ee
In obtaining \eq{a1} from \eq{a}, no approximation is used for the background so it is valid both during inflation and reheating. Because $\zeta$ is ill defined only at isolated times when $\dot{\f}=0$, one may still insist on using it in loop calculations since loops involve the time integrals of the $\zeta$-propagator and these may become  well defined although the propagator is singular. In that case, the superhorizon zeta modes can be found as 
\be\label{zh}
\zeta_k\simeq \zo+c_k f(t),
\ee
where $f$ is the same function defined in \eq{df}. Then, the singular $H^2/\dot{\f}^2$  terms coming from the $\z$ propagator can be seen to be canceled out by the $\df$ terms coming from the interaction vertices that arise by replacing $\vf$ with $\z$  using \eq{tr}. Repeating the loop calculations in the $\vf=0$ gauge, one may then check that the results for the corresponding graphs are simply the same with \eq{r1} and \eq{1b2}. For example, using the first order solution for the lapse $N=1+\dot{\zeta}/H$, the expansion of the  $\int\sqrt{h}NV$ term in the action gives the following interaction Hamiltonian after an integrating by parts
\be\label{iz1}
H_I(t)=-g^2\, a(t)^3\,\f(t)\,\int \, d^3x\,  \fr{\dot{\f}}{H}\,\z(t,\vec{x})\chi(t,\vec{x})^2,
\ee
which corresponds to \eq{i2}. The contribution of the 1-loop graph coming from this interaction to the $\zeta$-$\zeta$ correlation function can be seen to be equal to \eq{1b2}. As noted, the ${\dot{\f}}/H$ factor multiplying $\zeta$ in \eq{iz1} cancels out the singular term in the propagator. 

\section{The entropy perturbation and Weinberg's solution} 

As pointed out in the Introduction, in the presence of entropy perturbations the curvature perturbation $\zeta$ is not necessarily conserved. In a generic two-field model, \cite{ent} shows that while the adiabatic field $\s$ is defined as 
\be
\dot{\sigma}=(\cos\th)\df+(\sin\th)\dot{\chi},
\ee
the entropy perturbation is given by
\be
\d s=(\cos\th)\d\chi-(\sin\th)\d \f,
\ee
where
\be
\cos\th=\fr{\df}{\sqrt{\df^2+\dot{\chi}^2}},\hs{3} \sin\th=\fr{\dot{\chi}}{\sqrt{\df^2+\dot{\chi}^2}}. 
\ee
In our case, since we compute loop corrections till the end of the coherent inflaton oscillations, the backreaction of the $\chi$ field is negligible, and thus the background value of $\chi$ is zero. Therefore, we have $\s=\f$ and $\d s=\d \chi$, which shows that $\f$ is the adiabatic mode and $\chi$ is the entropy mode. We thus see an example of the entropy perturbation affecting the adiabatic mode, but not in the usual way, i.e by the real $\chi$ particles created out of the background, but through {\it virtual}  $\chi$ modes circulating in the loops. 

On the other hand, in \cite{wt} Weinberg showed  that  {\it as long as long as loop graphs are negligible},  nonlinear classical equations for metric and general matter perturbations have a ``constant" solution for superhorizon modes that coincides with the solution describing the metric and matter produced by the single-field inflation. Assuming that the results are valid in quantum theory in the Heisenberg picture, one may then use the classical field equations to read the sum of the tree graphs for a given quantum correlation function.   
Although, this result justifies the use of inflationary correlation functions to compute, for instance, the CMB fluctuations, it is emphasized in \cite{wt} that quantum fluctuations of large momenta circulating in the loops may invalidate it. As pointed out by Weinberg, generically the tree graphs make larger contributions and loops are  negligible. Our explicit computation shows a counterexample to this expectation due to the parametric resonance effects. 

\section{A Numerical estimate} \label{ap3}

In this appendix, we integrate \eq{r1} numerically for $g=10^{-2}$ and $m=10^{-6}\mt$. As shown in \cite{reh4}, 
and as it is pointed out several times above, at the end of the first stage of preheating one has $H\simeq10^{-2}m$. Thus, $mt_f\simeq 100Ht_f=200/3$, where we implement \eq{ha}. Since the first stage of preheating ends about 11 oscillations, one finds $mt_R\simeq 1$. From \eq{b1} the background inflaton field  can be fixed as
\be
\f(t)\simeq \fr{\mt}{20 m t}\sin(mt).
\ee
Using the maximum of $\lf\chi^2\rg$ given in \eq{cmax} and noting that $\lf\chi^2\rg$ increases exponentially with the effective index $\m$, one may obtain
\be
 \lf\chi^2(t)\rg\simeq \fr{m^2}{g^2}\exp(2\m m(t-t_f)).
\ee
For our case $\m\simeq0.13$. Using all this information, one can now integrate \eq{r1} numerically. We use Mathematica to evaluate this simple numerical integral that gives $P_{k\,\,\textrm{num}}^{\z(1)} \simeq  3.38\,P_k^{\z(0)}$ (see also footnote 5). 

\begin{acknowledgments}
I would like to thank Robert Brandenberger for useful discussions. 
\end{acknowledgments}

\end{document}